\documentclass[12pt]{iopart}

\usepackage[utf8]{inputenc}
\usepackage[T1]{fontenc}
\usepackage{amsmath2}
\usepackage{amssymb}
\usepackage{graphicx}
\usepackage{hyperref}
\usepackage{color}

\usepackage{algorithm}
\usepackage{algorithmicx}
\usepackage{algpseudocode}

\begin{document}

\title{Inferring financial stock returns correlation from complex network analysis}

\author{Ixandra Achitouv}
\ead{ixandra.achitouv@cnrs.fr}
\address{Institut des Syst\`emes Complexes ISC-PIF , CNRS, 113 rue Nationale,
Paris, 75013, France.}

\date{\today}

\begin{abstract}
Financial stock returns correlations have been studied in the prism of random matrix theory, to distinguish the signal from the "noise".  Eigenvalues of the matrix that are above the rescaled Marchenko–Pastur distribution can be interpreted as collective modes behavior while the modes under are usually considered as noise. In this analysis we use complex network analysis to simulate the "noise" and the "market" component of the return correlations, by introducing some meaningful correlations in simulated geometric Brownian motion for the stocks. We find that the returns correlation matrix is dominated by stocks with high eigenvector centrality and clustering found in the network. 
We then use simulated "market" random walks to build an optimal portfolio and find that the overall return performs better than using the historical mean-variance data, up to $50\%$ on short time scale.

\end{abstract}


\section{Introduction}
\label{introduction}
Analysing the eigenvalues distribution of stock returns correlation matrix provides insights into the complex structure of financial markets. By comparing empirical correlation matrices with those predicted by Random Matrix Theory (RMT), one can identify significant deviations that highlight genuine market correlations versus pure noise or noise mixed with "small" signal, see \cite{laloux1999noise}, \cite{plerou1999universal} for the first comparison of stock returns correlation matrix with RMT. This approach aids in improving portfolio optimization and risk management by filtering out noise \cite{laloux2000random}. Additionally, RMT analysis suggest the existence of collective behavior of stocks \cite{Akemann2010}, \cite{Guhr2003}, the market mode (which correspond to the highest eigenvalues) and other collective modes that could be sector-specific or follow other unknown economic forces. The interpretation of these collective modes is not straightforward from RMT analysis alone.

In this work we will show that complex science network applied to financial stock returns analysis can help interpret these collective modes. Indeed, complex science network analysis applied to financial stock has been used in various applications such as portfolio optimization, systemic risk assessment, understanding the diffusion of financial crisis. The structure of the network was correlated to the dynamics of the market on future fluctuation changes, crisis, stock returns, see \cite{heshmati2021portfolio} Tab1 where the authors report key articles on stock market network analysis from 1999-2021. 

In this work we combine the RMT analysis with the network analysis of the S$\&P$500 stocks to better understand the collective modes present in the stock returns correlations. We generate correlated random walks and compute the correlation of their returns using properties of the network and find that the resulting correlation of the returns captures what we observe in the data which allow us to identify cluster of stocks that influence each others. As an application we test how a Markowitz portfolio \cite{Markowitz1952} based on "market" modes increase the overall return.

\section{Construction of the network}

\subsection{Datasets}
In what follows we consider the stocks from the S$\&$P500 from 01-01-2019 to  01-01-2024 downloaded from Yahoo finance. 
We clear out stocks that were not present in the entire time range, ending up with 485 stocks and 1258 days of closing values for each stock. 

We also attribute a sector for each stock, scrapped from \url{https://en.wikipedia.org/wiki/List_of_S%26P_500_companies}. There are 11 sectors in total: Financials, Real Estate, Energy, Consumer Discretionary, Utilities, Communication Services, Health Care, Consumer Staples, Information Technology and Industrials, Materials.

\subsection{Network construction}
The structure of the financial market can be represented as a network where nodes represent financial entities such as stocks and the edges connecting them represent the correlations between their returns after applying some filtering e.g. \cite{boginski2005statistical}, \cite{onnela2004clustering}, \cite{kim2002weighted}, \cite{mantegna1999hierarchical}. 

\begin{figure}
    \centering
    \includegraphics[width=1.\linewidth]{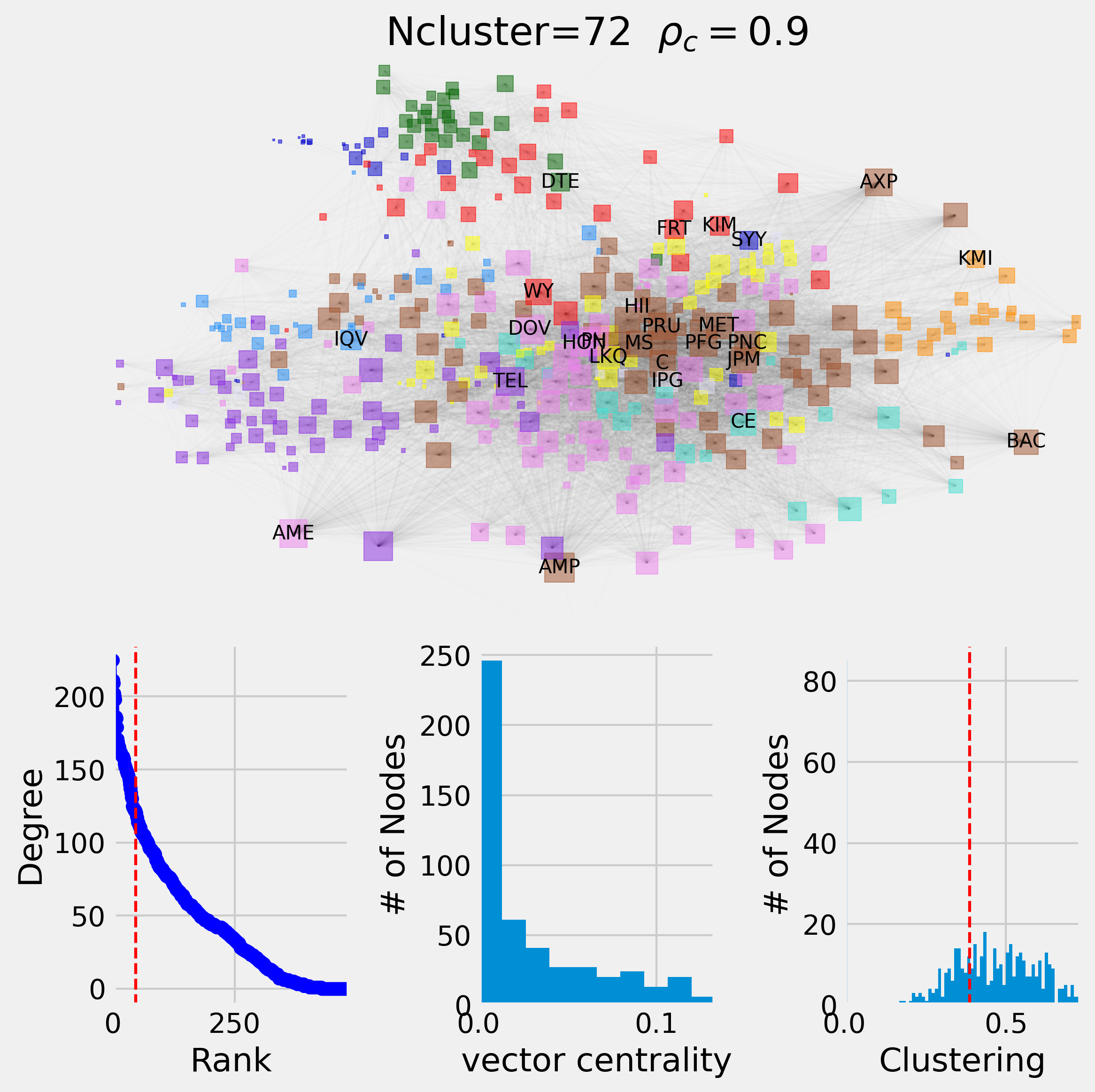}
    \caption{Network of the stocks (top panel) along with the degree distribution (left bottom panel), eigenvector centrality histogram (middle bottom panel) and clustering histogram of the stocks (right bottom panel). Red vertical lines correspond to the mean.}
    \label{fig:network}
\end{figure}

Thus we compute the correlation matrix of their closing price $P_i(t)$ log-return:
\begin{equation}
    C_{i,j}=\frac{< r_i(t) r_j(t)>-<r_i(t)><r_j(t)>}{\sigma_i\sigma_j}\label{eqCij}
\end{equation}
where $r_i(t)=\log[P_i(t)]-\log[P_i(t-1)]$ and $\sigma_i$ is the standard deviation of the stock computed over the period we consider. 

In this correlation matrix all stocks are connected to one another. One approach to convert this matrix to an adjacency matrix $A_{i,j}$ is the threshold method originally introduced in \cite{boginski2005statistical} and used in many studies e.g. \cite{namaki2011network}, \cite{huang2009cumulative}, \cite{xu2018efficient} where $A_{i,j}=C_{i,j}$ if $|C_{i,j}| \ge \rho_{c}$, otherwise $A_{i,j}=0$. The value for the threshold $\rho_c$ is usually an heuristic choice \cite{park_perspective_2020}. The lower the value the more connected are the stocks leading to a centered clustering distribution and a flat degree distribution with no preferential attachment. On the other other hand, a high value of the threshold leads to the well-known scale free network where the degree distribution follows a power law. This means that a few nodes are connected with many (high degree) while most are not connected (null degree). For instance in \cite{nobi2014effects}, $\rho_c=<C_{i,j}>+n\sigma$, where $\sigma$ is the standard deviation of the $C_{i,j}$ distribution and n an integer\footnote{Recently, a criterion based on network properties was introduce in \cite{Xu_2018}}. The higher the threshold the more we remove spurious (noise) correlations. In what follows we consider $\rho_c=0.9$, leading to a scale free network with 428 nodes and 11519 edges.

\subsection{Network analysis}
Once we have built the adjacency matrix, we use the NetworkX library \cite{hagberg2008exploring} to visualize the network and compute useful statistics.

\begin{itemize}
    \item The degree of a node: it is defined as the number of connections or edges it has to other nodes. 
    \item The eigenvector centrality: it measures the influence of a node in a network. A high eigenvector score means that a node is connected to many nodes who themselves have high scores. The eigenvector centrality for node i is
    \begin{equation}
\mathbf{Ax} = \lambda \mathbf{x}
    \end{equation}
where A is the adjacency matrix of the graph G with eigenvalue $\lambda$ \cite{newman2010networks}.
    
    \item Local Clustering coefficient: it is a measure of the density of connections between the node's neighbors defined as:
    \begin{equation}
   C_{i}=2\frac {|\{e_{jk}:v_{j},v_{k}\in N_{i},e_{jk}\in E\}|}{k_{i}(k_{i}-1)}, 
\end{equation}
where the numerator is the cardinal of the node $v_i$ neighbourhood (immediate connected neighbours), while $k_i(k_i-1)/2$ is the number of links that could possibly exist between node $v_i$ and its neighbours $k_i$. The global clustering coefficient is an average of the local clustering coefficients of all nodes in the network.
\end{itemize}

In Fig.\ref{fig:network} top panel, we display the resulting network, where nodes are colored by the sector of the stocks. The spatial visualisation of the network is computed using ForceAtlas2 \cite{jacomy2014forceatlas2} which maximizes/minimizes the distance of nodes that have low/high weighted edges respectively. The size of the nodes is proportional to their degree. Interestingly we can observe some clustering where stocks of the same sector (shown by colors) are closer. This is the case for the Utilities (green), the Energy (orange), the Information Technology (purple), the Health Care lightblue (lightblue), the Consumer Staples (dark blue) and the Real Estate (red), although some of their nodes are more densely connected to other sectors. On the other hand, the Financials (brown) nodes, the Consumer Discretionary (yellows), the Industrials  (pinks), the Materials cyan (cyan) and the Communication Services (whites) are quite spread across the network.

We run a Louvain community finder \cite{blondel2008fast} which optimizes locally  the difference between the number of edges between nodes in a community and the expected number of such edges in a random graph with the same degree sequence. 
As a result, we identify 72 clusters out of which 15 have more than one stocks.

For visibility we display the label of the nodes if the node is in the top 3$\%$ of the eigenvector centrality distribution, when its eigenvector centrality is the maximum within the 15 clusters, or when it has the largest eigenvector centrality within its S\&P5500 sector. 

In the lower panel of Fig.\ref{fig:network} we display the degree distribution of the nodes as well as the histograms of eigenvector centrality and local clustering. As expected, we have a scale-free network for the degree distribution. 
High eigenvector values can be problematic for systemic risks as contagion risk is linked to these influencial stocks \cite{zhu2016measuring}. Indeed, a highly interconnected financial network can act as an amplification mechanism, by creating channels for
a shock to spread, leading to losses that are much larger than the initial changes \cite{jackson2020systemicriskfinancialnetworks}. The red vertical line corresponds to the mean. Networks with high clustering coefficients are often more robust and resilient to node failures, as the redundancy in connections within clusters can help maintain the overall connectivity \cite{albert2000error}. This could be used as a network systemic risk measure similarly to the eigenvector centrality, when building on optimal portfolio. In fact it was found (e.g. \cite{heshmati2021portfolio}) that one can improve on traditional portfolio  selection by maximizing the usual Sharpe ratio where the standard deviation of the portfolio is replaced by a weighted average of node centrality measures (e.g. eigenvector centrality, betweennness centrality, closeness centrality).

\section{Models to capture the data correlation of stock returns}
In the seminal work of \cite{laloux1999noise}, \cite{plerou1999universal}, the eigenvalues distribution of the correlation matrix of stock returns is compared to the expected distribution of a pure uncorrelated matrix. For the period T (in days) where the stock returns are computed,  and the number of stocks N, such that N$\gg 1$, T$\gg 1$ and $T/N$ constant, the expected distribution of a pure random matrix follows a Marchenko-Pastur (MP)\cite{marchenko_pastur_1967} distribution:
\begin{equation}
    \rho(\lambda)=\frac{Q}{2\pi\sigma^2}\frac{\sqrt{(\lambda_{max}-\lambda)(\lambda-\lambda_{min})}}{\lambda}
\end{equation}
where $\lambda_{min},\lambda_{max}=\sigma^2(1+Q\pm 2\sqrt{1/Q})$, Q=T/N and $\sigma^2$ is the variance of the matrix, in our case equal to one.

The key results in \cite{laloux1999noise} are firstly: the maximum of the eigenvalues $\lambda_{market}$ is approximately an order of magnitude larger than $\lambda_{max}$. Secondly, by re-scaling $\sigma^2\rightarrow 1-\lambda_{market}$ or by treating $\sigma^2$ as an effective parameter, the continuous part of the distribution can be approximately fitted by a MP distribution (accounting for 94$\%$ of the spectrum \cite{laloux1999noise}), the other eigenvalues exceeding the continuous spectrum. The largest eigenvalue is related to a strongly localized eigenvector that presents the collective evolution of the system, and is therefore called the market mode. Its magnitude can be interpreted as the coupling strength of the system. In \cite{podobnik2010epl} it was shown that the time evolution of $\lambda_{market}$ tends to have peaks in times of crisis. The other exceeding values of $\lambda$ can also be interpreted as collective behaviours while the values following the MP distribution can be interpreted as spurious correlations among stocks. However we find that the distribution of the eigenvalues for the S\&P500 is nowadays quite different from the MP distribution in the continuum spectrum (see Fig1 of \cite{laloux1999noise}) with the one we compute in Fig.\ref{fig:spectra} left panel. The blue curve corresponds to the MP distribution and the red curve to the MP with $\sigma^2$ re-scaled. The red curve does capture the limit of the continuous spectrum but within the continuous spectrum the eigenvalues do not fill the contour of the curve. In \cite{laloux1999noise} the authors mention that a better fit could be obtained by allowing for a slightly smaller effective value of Q, which could account for the existence of volatility correlations. Indeed, if the degree of freedom within the continuous spectra is smaller than the number of modes then it is equivalent to having a smaller number of stocks, hence $Q$ should be smaller. Of course it does raise the question of interpreting "noise" if it contains correlations. 

In what follows, we introduce a model to capture the different components of the eigenvalues distribution, that we refer to as  "noise" and "market modes", with the aim to explain the qualitative features of the return correlation matrix. To this purpose we generate Geometrical Brownian Motion (GBM) that model the behaviour of the stocks closing price with some correlations among them. The correlations are modeled by hyper-parameters that are computed from the network properties.

\subsection{Introducing correlation in GBM}

The GBM model assumes that stock prices are log-normally distributed. 
In its integral form, the GBM model is used to describe the evolution of stock prices over time and is given by

\begin{equation}
    S_t = S_0 \exp \left( \left( \mu - \frac{\sigma^2}{2} \right) t + \sigma W_t \right)\label{eqGBM}
\end{equation}

where $S_t$ represents the stock price at time $t$, $S_0$ is the initial stock price, $\mu$ is the drift coefficient, representing the expected return of the stock, $\sigma$ is the volatility coefficient, representing the standard deviation of the stock's returns and $dW_t$ is a Wiener process (or Brownian motion), representing the random component of stock price changes. One of the key limitations is the assumption of constant volatility and drift, which fails to account for the changing economic conditions and market dynamics, leading to inaccuracies in modeling stock prices \cite{hull2014options}. Furthermore, the log-normal assumption fails to capture the heavy tails and skewness observed in actual stock return distributions. Therefore, the GBM model is useful although it can be improve using more sophisticate models that capture sudden large movements in stock prices, time-varying volatility, volatility clustering or mean-reverting behavior, observed in financial time series. In what follows we will show however that this simple toy model can capture the main features of the "signal" and "noise" we observe in the stocks returns correlation function.

To introduce correlations in our stocks we build a function with two parameters: the coefficient of the correlation $c_{\rm eff}\in[-1,1]$ and the common seed to generate the random walk: $s_{i}$, which is an integer. In eq.\ref{eqGBM}, the Wienne process is generated from 
$W_{t_j} =  \sqrt{dt}\sum_{i=0}^{i=j} \epsilon_i$ where $\epsilon_i$ is no longer a random value drawn from a normal (Gaussian) distribution of variance 1 and mean 0 but the contribution of two random values drawn from a normal distribution $\epsilon^{1},\epsilon^{2}$:
\begin{equation}
    \epsilon_i=\epsilon^{1}_{i}(1-c_{\rm eff})+c_{\rm eff}\epsilon^{2}_{i}
\end{equation}
where the seed used to generate $\epsilon^{2}=[\epsilon^{2}_{i=0},...\epsilon^{2}_{i=N}]$ is given by $s_{i}$ while the seed for $\epsilon^{1}$ is generated randomly for each stock. 

The parameters $S_0$, $\mu$ and $\sigma$ are computed from the historical data for every stocks. The pseudo code use to generate this model is written in appendix \ref{Pseudo_code}.

Hence, to compute simulated correlation matrix (for the stocks returns, the market or the noise returns correlations), we will generate a GBM for each of our S$\&$P 500 stocks and then compute eq.\ref{eqCij} from the resulting simulated walks.

\subsection{The market modes}

Rather then selecting the highest eigenvalue as the market mode in the distribution of the eigenvalues, we consider multiple market modes by selecting the largest n values of the eigenvalue distribution. We select n such that it equals the number of the most "influential" stocks in the network. To select the most influential stocks we require that both the eigenvector centrality and their pagerank are in the top 3$\%$ of the distributions. Then we check the smallest market eigenvalue is $>=\lambda_{MP}$ where $\lambda_{MP}$ is the expected largest mode in the MP distribution (without any rescaling). 
We end up with $n_{market}=12$ stocks that are in the Financial sector (7),  Industrial sector (4) and Information Technology sector(1). Their labels are [AME, AMP, AXP, C, DOV, HON, MET, MS, PFG, PH, PRU, TEL]. 

The largest 12 eigenvalues of the spectral distribution are $\lambda^{market}=$ 
[203.87,
27.13,
21.99,
9.88,
7.63,
5.7,
5.41,
4.79,
4.13,
3.92,
3.82,
3.39,
3.08]
These eigenvalues are assumed to be the market modes and will be use to build the market correlation matrix of the returns. 

We assume that all the other eigenvalues are part of the "noise modes" and will be use to build the noise correlation matrix of the returns. 

\subsection{Extracting Market correlation returns from the data.}

To build the market correlation matrix of the returns M from the data we perform an eigendecomposition on the $n_{market}=12$ eigenvalues selected, such that 
\begin{equation}
    M=Q\Lambda Q^{-1}\label{eqED}
\end{equation}
where Q is the square 
Nstock$\times$ Nstock matrix whose i column is the eigenvector 
$qi$ of $C_{i,j}$ and $\Lambda$ is the diagonal matrix whose diagonal elements are the corresponding eigenvalues, $\Lambda_{ii} = \lambda^{market}_{i}$ for $i\in [1,n_{market}]$, and  $\Lambda_{ii} =0$ for $i\in [n_{market+1},\rm Nstocks]$. We then apply a rescaling such that we define 
\begin{equation}
    C_{i,j}^{Market}=\frac{M_{i,j}}{\sqrt{M_{i,i}M_{j,j}}}, 
\end{equation}

the correlation matrix that contains the market modes (or collective modes) that emerge from a system with many degrees of freedom (Nstocks). 

In order to understand how the different stocks interact to form these collectives modes, we try to reproduce this matrix by generating correlated random walks.

\subsection{Simulating Market correlation returns from correlated GBM}\label{labsecmarketsim}

To generate the market modes we introduce two types of correlation:
\begin{itemize}
    \item Correlations per Louvain communities  
    
For the correlation per Louvain communities, the seed $s_i$ is given by the label of the community cluster the stock belong to thus it is a fixed integer which correspond to one of the 72 clusters. The strength of the correlation $c_{\rm eff}$ is given by the clustering value of the stock in the network. In our model the clustering value is defined as the geometric average of the subgraph edge weights\cite{onnela2005intensity}. 

    \item Correlations per Market modes

For the correlations per market modes, the seed $s_i=4376$ is arbitrary and fixed for all stocks. The strength of the correlation $c_{\rm eff}=C_{is,jm}$ where $is$ corresponds to the stock we consider and $jm$ to the stock within our market mode list for which $|C_{is,jm}|$ is the largest.  

\end{itemize}
The interpretation of these two terms is quite intuitive. When a stock is within a cluster, not given by its sector but from the network properties, its evolution couples to the other nodes within this cluster, proportionally to its degree of clustering. The correlations per market modes capture the global trend of the stock to follow the trend set by one of the highly influential stocks, the one with the strongest connection. 

Then for each stock we simulate a walk with correlations per Louvain communities: $S^{L}_{t}$ and one with correlations per Market modes $S^{M}_{t}$. The final walk is a weighted contribution of the two:
\begin{equation}
    S_{t}=\frac{w_L S^{L}_{t}+w_M S^{M}_{t}}{w_L+w_M}
\end{equation}
with $w_L+w_M=1$.

Once we simulate Market GBM for each of the S\&P500 stocks we can compute the correlation of the returns for these walks as per eq.\ref{eqCij} leading to a simulated market correlation of the returns: $C_{i,j}^{Market\; GBM}$. 

The weights $w_L, w_M$ are the same for each stocks and they are
found by running a simple stochastic search which minimizes the Wasserstein distance (e.g. \cite{villani2008optimal}) between the distribution of $C_{i,j}^{Market}$ and $C_{i,j}^{Market \; GBM}$.

\subsection{Extracting noise correlation returns from the data}

To extract the "noise" modes from the data we select all of the eigenvalues that are not in $\lambda^{market}$ and perform the same eigendecomposition of eq.\ref{eqED} with $\Lambda_{ii}=0$ for $i\in[1,n_{market}]$, leading to a correlation matrix of the "noise" returns $C_{i,j}^{noise}$.

\subsection{Simulating noise correlation returns from GBM}
To generate the noise mode we could consider solely uncorrelated GBM. However as we mentioned previously, it seems a smaller value of $Q$ would provide a better fit to the MP distribution suggesting some correlation within the continuous spectrum. Hence we simulate the noise walk for each stock as a contribution of two terms:
\begin{itemize}
    \item A pure GBM walk $S_{t}^{N}$, where the seed $s_i$ is irrelevant as we set $c_{\rm eff}=0$ for each walk. 
    \item Correlations per Louvain communities as explained in sec.\ref{labsecmarketsim}
\end{itemize}

The final walk is given by:
\begin{equation}
    S_{t}=\frac{w_L S^{L}_{t}+w_N S^{N}_{t}}{w_L+w_N}
\end{equation}
 
with $w_L+w_N=1$. 

Once we simulate "noise" GBM for each of the S\&P500 stocks we can compute the correlation of the returns for these walks as per eq.\ref{eqCij} leading to a simulated noise correlation of the returns: $C_{i,j}^{Noise \; GBM}$. 

The optimal weight are the same for each stock and are found using the same stochastic search, minimizing the Wasserstein distance between the distribution of $C^{noise}_{i,j}$ with $C_{i,j}^{Noise \; GBM}$.   

\subsection{Simulating the total correlation returns from  correlated GBM}

To simulate the correlation function of the returns we use for each walk a weighted contribution of the previous three models:

\begin{equation}
    S_{t}=\frac{w_L S^{L}_{t}+w_M S^{M}_{t}+w_N S^{N}_{t}}{w_L+w_M+w_N}
\end{equation}
with $w_L+w_M+w_N=1$. Again we can compute the correlation of the returns for these walks as per eq.\ref{eqCij} leading to a simulated correlation of the returns: $C_{i,j}^{GBM}$.

We find the optimal weights by running a simple stochastic search on a two dimensional grid, whose parameters minimize the Wasserstein distance between the distribution of $C_{i,j}$ and $C_{i,j}^{GBM}$. 

Note that initially we tried to use the sector of the stock instead of the Louvain clustering, but the former fails to capture the trend of the market and the total correlation of the returns.

\section{Results}

\begin{figure}
    \centering
    \includegraphics[width=1.\linewidth]{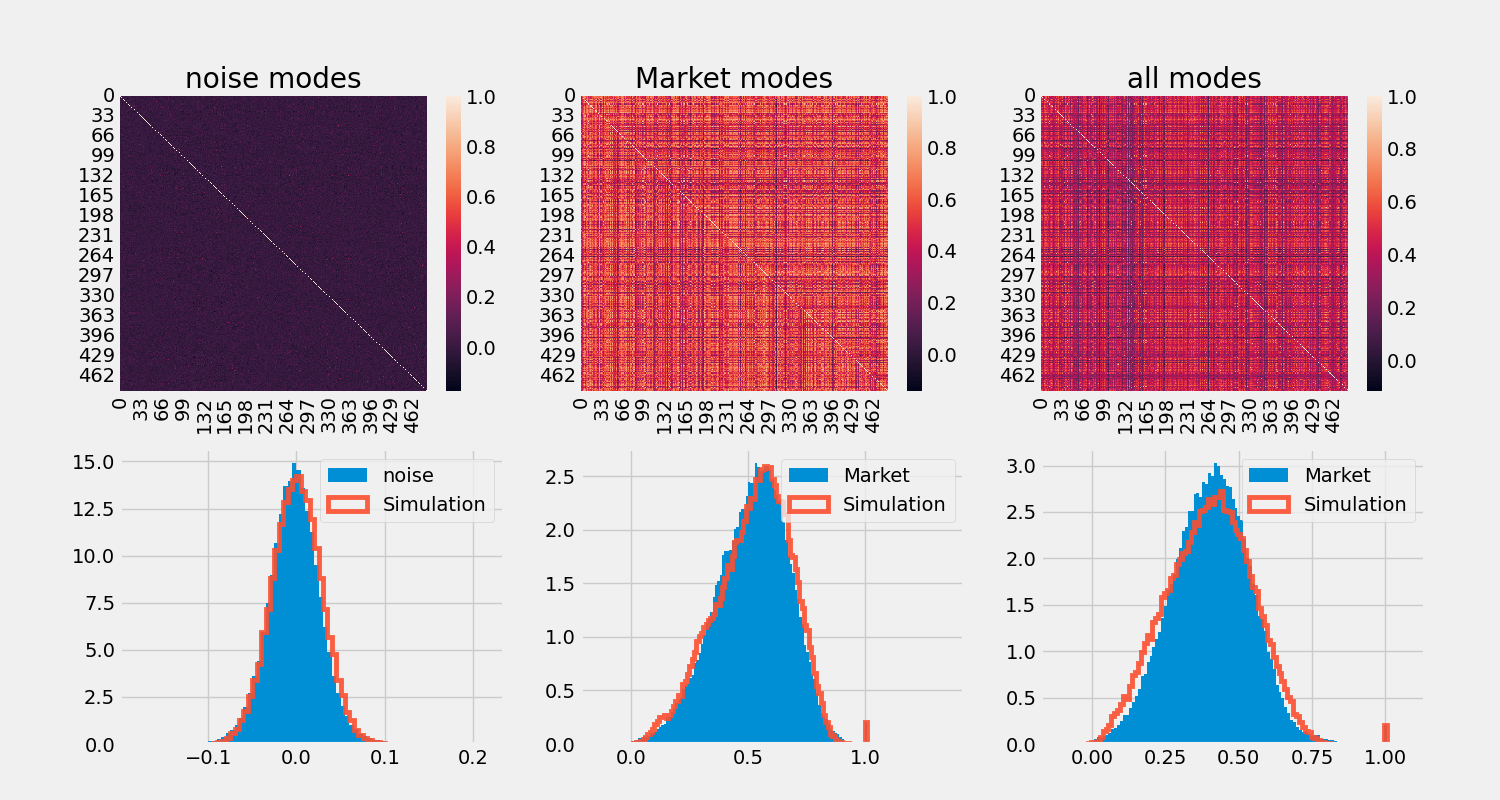}
    \caption{Top panel: decomposition of the signal (market modes), the noise and the complete correlation of the returns. Lower panel: distribution of the values in the data (blue histogram) and for the simulated walks (red histograms)}
    \label{fig:corr}
\end{figure}

\begin{figure}
    \centering
    \includegraphics[width=0.8\linewidth]{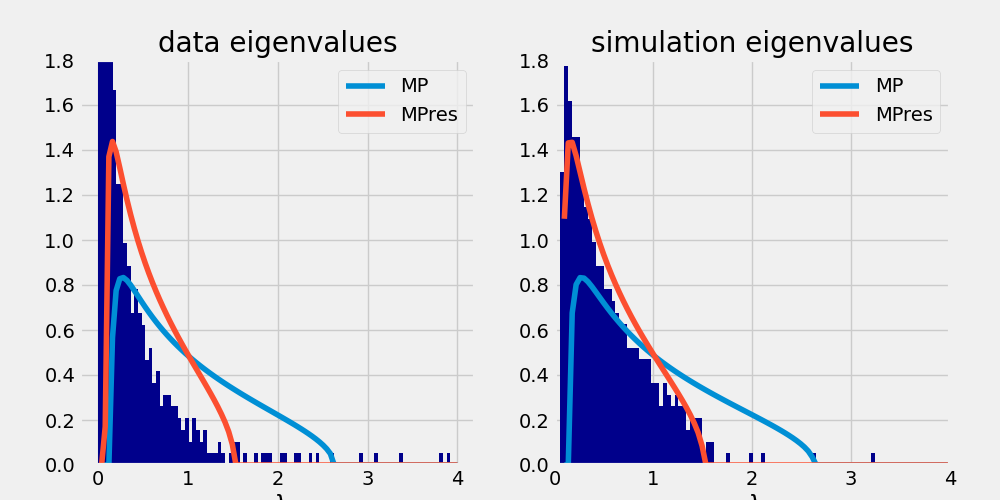}
    \caption{Distribution of the eigenvalues of the correlation function of the returns in the data (left panel) and in our simulation (right panel). The red/red curves correspond to the MP distribution rescaled/unrescaled respectively. }
    \label{fig:spectra}
\end{figure}

In Fig.\ref{fig:corr} we can see on the top panels the 3 correlation matrices computed from the data: $C^{noise}_{i,j}$, $C^{Market}_{i,j}$ and $C_{i,j}$.

In the lower panels we display the distribution of their components (blue histogram). The red histograms correspond to the results we obtain from our simulated walks: it is the distribution of $C_{i,j}^{Noise \; GBM}$, $C_{i,j}^{Market \; GBM}$ and $C_{i,j}^{GBM}$. 

First we observe that the noise modes are centered on zero with a standard deviation of $0.05$ as for the simulated noise walks. The market modes distribution has a mean and a standard deviation of 0.52, 0.16 for the data, and 0.50, 0.17 for the simulation. In the full correlation $C_{i,j}$ the mean and standard deviation are 0.41, 0.14 and 0.42, 0.15 in $C_{i,j}^{GBM}$. 
The simulated distributions do capture both the trend of the data and the split between noise and signal seem reasonable. 

Interestingly the best weights for the market modes are $(w_L=0.26; w_M=0.74)$ which suggests that $74\%$ of the returns can be captured by the 12 market stocks we identify using the network properties. The other correlated modes correspond to the 15 (greater than one) clusters. For the noise modes the weights are $(w_L=0.03;w_N=0.97)$, suggesting that the noise modes are well captured by a pure GBM and that the lower value of $Q$ should be model by some different type of correlation such as volatility correlations as suggested in \cite{laloux1999noise}. For the total correlation function of the returns, the weights are $w_{N}=0.18, w_L=0.20; w_M=0.62)$ suggesting that $18\%$ of the returns correlation corresponds to spurious correlations while the other $82\%$ are driven by the market stocks and the cluster the stock belongs to.

In Fig.\ref{fig:spectra} we show the eigenvalue distribution of $C_{i,j}$ (left) and the simulation $C_{i,j}^{GBM}$. The blue and red curves correspond to the rescaled and unrescaled MP prediction. The largest eigenvalues of the two distributions are similar hence are the red curves: 203.9,205.4 for the data, simulations respectively. With our models we have a few decoupled modes from the continuous spectrum, but not all. The distribution within the continuous spectrum is less convex than in the data. Again we expect it to be the case due to the simple GBM model we used that does not capture correlation in the volatility \cite{laloux1999noise}. The limit of the continuous spectrum however is approximately the same, around $1.6$. 

Overall we find that a very simple model (GBM with correlations) can explain the overall interactions between the stocks leading to an agreement of the qualitative features of the return correlation matrix. The splitting of the market modes correlations provide a better understanding of the collective modes which can be explained by a few market modes and a few clustering among the stocks. This provides an extended perspective on the collective modes compared to the RMT analysis. 

\section{Application: construction of an optimal portfolio using Market GBM}

In the seminal work of Markowitz \cite{Markowitz1952}, the construction of a portfolio aims to minimize risk for a given level of expected return by considering the correlation between asset returns. By combining assets that are not perfectly correlated, investors can reduce overall portfolio risk and achieve a more efficient risk-return trade-off. In this Mean-Variance Optimization model, the risk of a portfolio is defined as the variance of the portfolio $\sigma_{P}^{2}=\sum_{i,j=1}^{N} w_i Cov_{i,j} w_j$, where $w_i,w_j$ are the weights of the assets  $i,j$ and $Cov$ the covariance matrix among the N assets. The average return of the portfolio is $R_P=\sum_{i=1}^{N} w_i R_i$, where $R_i$ is the expected return of asset $i$. Finding the optimal weights thus maps into solving a quadratic optimization problem:

\begin{align}
    &\min_{\mathbf{w}} \quad \mathbf{w}^\top Cov \;\mathbf{w} \\
    &\text{subject to:} \quad \mathbf{w}^\top \mathbf{R} = R_p \\
    &\qquad \qquad \sum_{i=1}^N w_i = 1 ,  w_i \geq 0 \quad \forall i 
\end{align}

with $\mathbf{w}$ is the vector of portfolio weights and $\mathbf{R}$ is the vector of expected returns for each asset. This can be found using a Lagrange multiplier \cite{BUN20171}, assuming that $Cov$ can be inverted. 

In this approach, the expected return of each stock and the covariance among stocks are unknown. One crude approximation is to use historical returns of the stocks and compute their covariance, though more reliable models \cite{fabozzi2010robust} have been proposed to estimate the expected returns (e.g. \cite{black1992global}) or the covariance matrix of the returns \cite{ledoit2003improved} -- it is still an active field of research. In what follows, we consider as a baseline scenario the historical values of the stocks to compute the Mean-Variance Optimization model and we select the weights that maximize the Sharpe ratio $S=(R_P-R_f)/\sigma_P$, using the {\it PyPortfolioOpt} software \cite{Martin2021}. We also add the constraint that the weights should be $\ge 0.0005$, to build a portfolio with all the stocks rather than a few selected ones. 

In comparison we want to test if the simulated Market GBM can be used to improve the overall return of the baseline portfolio. Indeed, we expect that the correlations among stocks are highlighted without the noise contribution. Hence we can expect to get a higher return. To test this hypothesis, we generate simulated Market GBM (defined in \ref{labsecmarketsim}) for each stock, from which we compute the simulated returns and their covariance. We use these two inputs to solve the quadratic optimization problem similarly to the baseline scenario.

\begin{figure}
    \centering
    \includegraphics[width=0.6\linewidth]{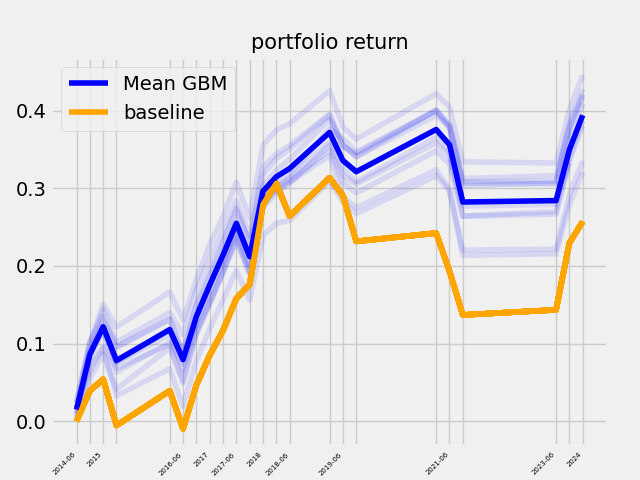}
    \caption{Portfolio return computed with the baseline scenario (orange curve) and for the simulated market correlated GBM (blue light curves). The blue curve corresponds to the mean of the 10 simulated walks.}
    \label{figport}
\end{figure}

We consider a 10-year period from 01-06-2014 to 01-06-2024. In this period we re-compute the weights on different time scales dT=[252; 252/2; 252/3; 252/4; 252/6; 252/8] (252 days corresponds to business days). For each time scale we solve for the optimal weights to compute the portfolio return on the next time scale. We generate 10 simulated random walks for each time scale to evaluate the variance of our predictions.  

We find that: 1--our model outperforms the baseline model for dT=[252/3; 252/4; 252/6; 252/8], while for larger time scales the baseline model provides more accurate returns. This can be explained by the fact that the generated random walks have mean returns that can be significantly different from the future mean returns of stocks for large time scales. 2--The optimal time scale to re-balance the portfolio corresponds to dT=253/3, rounded to 84 days. For this time scale the return of the baseline is $65\%$ and for the simulated walks it is $69\%$, a $6\%$ improvement. 3--The largest difference between the baseline model and the simulated walks occurs for dT=252/4. For this time scale we show in Fig.\ref{figport} the cumulative optimal portfolio return using the baseline scenario (orange curve) and the Mean-Variance model with our simulated Market random walks (blue curves).\footnote{Note that on some time scales the {\it PyPortfolioOpt} solver couldn't solve the quadratic optimization for the baseline scenario, leading to an x-axis with uneven tick-marks.} The cumulative return for the baseline scenario is $26\%$ while for the simulation it is $39\%$, a $50\%$ improvement. This is an interesting result, showing that the complex network analysis applied to extract the market mode can be use to improve the construction of an optimal portfolio with a different approach than in \cite{heshmati2021portfolio}.

\section{Summary and conclusions}
In this work we split the S$\&$P500 stock returns correlation into two components, market modes correlations and noise correlations. Using network science analysis tools we were able to generate a network of stocks which displays several interesting properties: a scale free network, clustering of stocks that is different from the stock sectors of the S$\&$P500, and key stocks that are highly influential.

The properties of the network were used to generate random walk correlations for the stocks. The properties of the correlations are: 1--the clustering coefficient of the nodes, 2--the eigenvectors centrality and, 3--the Louvain communities of the nodes. 

With these properties we simulate correlation matrices of the market modes return and for the full return correlation. We find that two sources of correlation (by market modes and by communities) can capture the main features of the observed stock returns correlations using a simple GBM for the walks, which is quite remarkable. 

We then use the simulated market GBM to build an optimal portfolio and find that it outperforms standard mean-variance models based on historical prices when we re-balance the weights every dT days, with dT$\le 84$, demonstrating the pertinence of using complex network science to extract the market modes.  

To conclude, this work provides a concrete interpretation of the collective modes observed in the distribution of eigenvalues of the financial stock returns correlations matrix. This allows us to identify clusters of stocks that behave in the same way (Louvain cluster) as well as key market players. We also show that the correlated GBM walks can be use for improving portfolio returns. Finally, this work could be expanded in several ways. More sophisticated models than the GBM could be employed, for instance one could implement a Heston model \cite{heston1993closed} to allow time-varying volatility, or a GARCH Model \cite{bollerslev1986generalized} to capture volatility clustering.

\section*{Data and code declaration}
The data used for this project can be downloaded from Yahoo finance's API. The code used to generate this analysis can be provided upon request and will be posted online after revision of this manuscript at \url{https://github.com/IxandraAchitouv}


\section*{Acknowledgements}
I would like to thank Maxime Viallet (Quantitative Analyst at G-Research) and Vincent Lahoche  (researcher at CEA) for their discussions and their comments upon the first version of this manuscript.


\section{Pseudo code for Simulate GBM Stock Price with Correlations}\label{Pseudo_code}
\begin{algorithm}
\caption{Simulate GBM Stock Price with Correlation}
\begin{algorithmic}
\Function{SimulateStockPriceGBMCorr}{$S0, mu, sigma, T, coef, seedi$}
    \Comment{
    Parameters:\\
    S0 (float): Initial stock price.\\
    mu (float): Drift coefficient (expected return).\\
    sigma (float): Volatility (standard deviation of returns).\\
    T (float): Total time period.\\
    dt (float): Time step.\\
    coef (float) : strength of the correlation\\
    seedi (integer) : si seed used for correlation\\
    Returns:\\
    np.array: A simulated stock prices\\
    }
    \State $dt \gets 1$ \Comment{Time step}
    \State $N \gets \lfloor T / dt \rfloor$ \Comment{Number of time steps}
    \State $t \gets$ Create a time grid from $0$ to $T$ with $N$ points
    \State Generate standard normal random variables $W1$ of size $N$ with random seed 
    \State Generate standard normal random variables $W2$ of size $N$ with seed= $seedi$
    \For{$i$ in range $0$ to $N-1$}
        \State $W[i] \gets W1[i] * (1 - coef) + coef * W2[i]$ \Comment{Generate correlated random numbers}
    \EndFor
    \State $W \gets$ Cumulative sum of $W$ multiplied by $\sqrt{dt}$ \Comment{Wiener process}
    \State $X \gets (mu - 0.5 * sigma^2) * t + sigma * W$ 
    \State $S \gets S0 * exp(X)$ \Comment{Final stock prices}
    \State \Return $S$
\EndFunction
\end{algorithmic}
\end{algorithm}


\bibliography{networkfinance.bib}


\bibliographystyle{iopart-num}

\end{document}